\documentclass[rnote,traditabstract]{aa}  
\usepackage{graphicx}
\usepackage{txfonts}
\usepackage{natbib}
\bibpunct{(}{)}{;}{a}{}{,} % to follow the A&A style
\begin{document}
   \title{Inelastic Na+H collision data for non-LTE applications in stellar atmospheres}

   \author{P. S. Barklem\inst{1}
          \and
          A. K. Belyaev\inst{2}
          \and
          A. S. Dickinson\inst{3}
          \and
          F. X. Gad\'ea\inst{4}
          }

   \institute{Department of Physics and Astronomy, Uppsala University, Box 515 S-75120 Uppsala, Sweden 
         \and
             Department of Theoretical Physics, Herzen University, St.\ Petersburg 191186, Russia
         \and
             School of Chemistry,  Newcastle  University, Newcastle upon Tyne NE1 7RU, United Kingdom
         \and
             Laboratoire de Chimie et Physique Quantique, UMR 5626 du CNRS, IRSAMC,
             Universit\'e Paul Sabatier, 118 rte de Narbonne, F-31062, Toulouse, France   
             }

   \date{Received 4 June 2010 ; accepted 25 June 2010}

% \abstract{}{}{}{}{} 
% 5 {} token are mandatory
 
  \abstract
  {Rate coefficients for inelastic Na+H collisions are calculated for all transitions between the ten levels up to and including the ionic state (ion-pair production), namely Na(3s,3p,4s,3d,4p,5s,4d,4f,5p)+H(1s) and Na$^+$+H$^-$. The calculations are based on recent full quantum scattering cross-section calculations.  The data are needed for non-LTE applications in cool astrophysical environments, especially cool stellar atmospheres, and are presented for a temperature range of 500--8000~K.  From consideration of the sensitivity of the cross-sections to input quantum chemical data and the results of different methods for the scattering calculations, a measure of the possible uncertainties in the rate coefficients is estimated. }


   \maketitle
%
%________________________________________________________________

\section{Introduction}

Lack of accurate data for collision processes needed for reliable non-LTE line formation calculations in cool star atmospheres, especially processes involving hydrogen atom and electron impacts, poses a major source of uncertainty for stellar abundance analyses; e.g. \citet{1993PhST...47..186L,2001NewAR..45..559K,2005ARA&A..43..481A}.  For hydrogen, the situation is particularly poor.  The possible importance of collisions with neutral hydrogen in non-LTE line formation calculations was first pointed out by \citet{1984A&A...130..319S} in their study of the of the statistical equilibrium of Li in cool stars.  Although inelastic processes due to hydrogen collisions are expected to be much less efficient than those due to electrons, the much greater abundance of hydrogen atoms may overcome this reduced efficiency: in the line forming regions of a solar-type star the abundance of hydrogen atoms is typically four orders of magnitude greater than that of electrons, even more in metal-poor stars. 

At the time of Steenbock and Holweger's study there was practically no reliable experimental or theoretical work on inelastic hydrogen collision processes; however, the situation has been improving slowly but steadily over the quarter of a century since then.  Their work prompted an experimental study by \citet{FGSSV:91} of $\mathrm{Na}(3s) + \mathrm{H} \rightarrow \mathrm{Na}(3p) + \mathrm{H}$ at low (15--1500 eV) energies, though due to experimental difficulties not down to near the threshold (2.1 eV for this case), which is the relevant regime for the temperatures of interest in cool stars.  Revised experimental data, including results down to 10~eV, were presented in \citet{BGHM:99}.  This work was followed by a number of theoretical studies involving some of the present authors.  First, quantum scattering calculations were performed for $\mathrm{Na}(3s) + \mathrm{H} \rightarrow \mathrm{Na}(3p,4s) + \mathrm{H}$ down to the threshold  \citep{BGHM:99} and found good agreement with the experimental results above 10 eV.   This work on Na+H was followed by calculations for Li+H \citep{2003PhRvA..68f2703B} based on quantum-chemical data calculated by some of us~\citep{CDG:99a}.  This was followed by astrophysical application \citep{2003A&A...409L...1B,2009A&A...503..541L}, where it was found that excitation collisions $\mathrm{Li}(nl) + \mathrm{H} \rightarrow \mathrm{Li}(n'l') + \mathrm{H}$ were unimportant, yet the ion-pair production and mutual-neutralisation process $\mathrm{Li}(3s) +\mathrm{H} \rightleftharpoons \mathrm{Li}^+ +\mathrm{H}^-$ (often referred to as charge exchange in astrophysics) was found to be rather important, resulting in changes in spectral line strengths of around 20\% in cool, metal-poor, sub-giant stars.

In a recent paper \citep{2010PhRvA..81c2706B} we revisited low-energy Na+H collisions, since for astrophysical non-LTE modelling of Na line formation data for transitions between all possible Na levels are needed, while the earlier experimental and theoretical studies dealt primarily with the resonance transition (which corresponds to the Na D lines).  Details of the calculations can be found in that paper.  Cross-sections for transitions between all ten levels up to and including the ionic state (ion-pair production) for collision energies from threshold to 10~eV were presented.  In fact, cross-sections were calculated up to collision energies of 100~eV; however, the results at energies higher than 10~eV are of little importance at the temperatures of interest.  The purpose of this research note is to present rate coefficients calculated from these cross-sections, as these rate coefficients are needed for non-LTE applications such as in cool stars.

\section{Collision rate coefficients}

The rate coefficients, $\langle \sigma v \rangle$, for excitation and deexcitation processes $\mathrm{Na}(nl) + \mathrm{H(1s)}  \rightleftharpoons  \mathrm{Na}(n'l') + \mathrm{H(1s)}$, and for the ion-pair production and mutual-neutralisation processes involving the ionic state $\mathrm{Na}(nl) + \mathrm{H(1s)} \rightleftharpoons \mathrm{Na}^+(2s^22p^6) + \mathrm{H}^- $, are presented in Table~\ref{tab:rates}.  The coefficients have been obtained by folding the cross-sections with a Maxwellian velocity distribution from threshold to 100 eV, and are presented for temperatures in the range 500--8000~K.  

The 500~K data are provided since there is substantial interest in Na lines in low temperature astrophysical environments such as brown dwarfs and planetary atmospheres \citep[e.g.][]{RevModPhys.73.719,2002ApJ...569L..51B}, and though other perturbers such as H$_2$ and He are usually more abundant in these situations, the data may be useful. \cite{PhysRevA.78.052706} have recently calculated data for inelastic processes in Na+He collisions with application to planetary and brown dwarf atmospheres in mind.  However, the main driver behind our study is the need for data for line formation modelling in F, G and K star atmospheres where ground state hydrogen atoms are indeed the most abundant perturber.  Thus, our discussion below will predominantly focus on the 2000--8000~K data.  

Although the data for inverse processes are related via the detailed-balance relation and thus redundant, little space can be saved by omitting one of the two and so we present data for both, thus saving the end user the need to make such calculations themselves.  It is worth noting that the excitation and ion-pair production processes were calculated directly from the cross-sections, while dexcitation and mutual-neutralisation processes were calculated via the detailed-balance relation rather than direct calculation from the deexcitation cross-sections.  As for Li \citep{2003A&A...409L...1B}, for the temperatures most relevant for cool stars around 6000~K, it is readily seen that the rate coefficients for ion-pair production are remarkably large, particularly from the first excited $s$ state, here 4$s$.  Some rates between close-together neighbouring levels, such as $4s \rightarrow 3d$ and $4d \rightarrow 4f$ are also seen to be very large.  At low temperatures, the rate coefficients for these transitions with small energy thresholds become by far the largest.

In \cite{2010PhRvA..81c2706B} it was shown that narrow orbiting and Feshbach resonances are often present in the cross-sections at low energy (see Figs.~5, 8 and 9 of that paper).  To fully resolve these features requires calculations with very small steps in collision energy, which is extremely time consuming.  Such detailed calculations were performed for the $3s\rightarrow 3p$ cross-sections in order to resolve a significant fraction of the orbiting resonances.  This further allowed pure background cross-sections, i.e. with no resonances, to be constructed for this transition.  From these calculations it was estimated that the contribution of the resonances to the rate coefficient was roughly 33\% at 500~K, 20\% at 2000~K, 15\% at 5000~K and 10\% at 8000~K.  The rate coefficients for $3s\rightarrow 3p$ presented here are calculated including the contribution of orbiting resonances.  In all other cases, the cross-section calculations were performed using an energy grid with steps much greater than the widths of the resonances.  Consequently most resonances are missed and those resonances that are by chance sampled are not resolved.  Missed resonances will lead to an underestimate of the contribution of the rate coefficient, while unresolved resoances will generally lead to an overestimate of the contribution since the widths of the resonances are significantly overestimated, and this may lead to some cancellation of errors.  In any case, the error in the rate coefficients from this source of uncertainty should not exceed the magnitude of the corrections given above for $3s\rightarrow 3p$, i.e. of order 10\%, which is significantly less than the uncertainties from other sources discussed below.

\begin{table*}
\caption{Rate coefficients $\langle \sigma v \rangle$, in units of cm$^3$~s$^{-1}$, for selected temperatures in the range $T=500$--8000 K, for the excitation and deexcitation processes Na($nl$)+H($1s$)$\rightarrow$Na($n^\prime l^\prime$)+H($1s$), or where indicated ion-pair production Na($nl$)+H($1s$)$\rightarrow$Na$^+$($2s^22p^6$)+H$^-$ and mutual neutralisation Na$^+$($2s^22p^6$)+H$^-\rightarrow$Na($n^\prime l^\prime$)+H($1s$).} 
\label{tab:rates}
%\scriptsize
\tiny
\begin{center}
\begin{tabular}{ccccccccccc}
\hline
Initial & \multicolumn{9}{c}{final state $n^\prime l^\prime$}  \\
state $nl$   & 3s & 3p & 4s & 3d & 4p & 5s & 4d & 4f & 5p & Na$^+$+H$^-$  \\
\hline
\multicolumn{11}{c}{\underline{ 500 K}} \\
          3s &       ---      & $ 1.07$E$-37$  & $ 1.09$E$-50$  & $ 3.41$E$-54$  & $ 8.22$E$-56$  & $ 2.75$E$-61$  & $ 8.05$E$-64$  & $ 2.76$E$-64$  & $ 9.36$E$-66$  & $ 5.64$E$-60$ \\
          3p & $ 5.81$E$-17$  &       ---      & $ 1.77$E$-23$  & $ 5.45$E$-28$  & $ 2.03$E$-29$  & $ 8.25$E$-36$  & $ 9.86$E$-40$  & $ 5.60$E$-40$  & $ 3.41$E$-43$  & $ 4.64$E$-34$ \\
          4s & $ 1.65$E$-18$  & $ 4.93$E$-12$  &       ---      & $ 4.15$E$-14$  & $ 7.63$E$-16$  & $ 2.77$E$-22$  & $ 2.93$E$-27$  & $ 2.61$E$-27$  & $ 6.14$E$-30$  & $ 4.14$E$-20$ \\
          3d & $ 2.02$E$-18$  & $ 5.95$E$-13$  & $ 1.62$E$-10$  &       ---      & $ 6.19$E$-13$  & $ 3.91$E$-20$  & $ 6.67$E$-24$  & $ 2.21$E$-24$  & $ 3.40$E$-27$  & $ 2.20$E$-17$ \\
          4p & $ 1.92$E$-18$  & $ 8.72$E$-13$  & $ 1.17$E$-10$  & $ 2.44$E$-11$  &       ---      & $ 1.42$E$-14$  & $ 7.98$E$-19$  & $ 4.21$E$-19$  & $ 2.20$E$-23$  & $ 2.53$E$-16$ \\
          5s & $ 8.97$E$-20$  & $ 4.96$E$-15$  & $ 5.96$E$-13$  & $ 2.15$E$-14$  & $ 1.99$E$-10$  &       ---      & $ 3.45$E$-15$  & $ 8.97$E$-16$  & $ 2.01$E$-18$  & $ 5.75$E$-15$ \\
          4d & $ 2.60$E$-21$  & $ 5.87$E$-18$  & $ 6.24$E$-17$  & $ 3.64$E$-17$  & $ 1.10$E$-13$  & $ 3.42$E$-14$  &       ---      & $ 6.18$E$-10$  & $ 1.83$E$-16$  & $ 8.69$E$-17$ \\
          4f & $ 7.34$E$-22$  & $ 2.75$E$-18$  & $ 4.57$E$-17$  & $ 9.91$E$-18$  & $ 4.80$E$-14$  & $ 7.32$E$-15$  & $ 5.09$E$-10$  &       ---      & $ 1.64$E$-16$  & $ 5.80$E$-17$ \\
          5p & $ 2.08$E$-22$  & $ 1.40$E$-20$  & $ 9.02$E$-19$  & $ 1.28$E$-19$  & $ 2.10$E$-17$  & $ 1.37$E$-16$  & $ 1.26$E$-15$  & $ 1.38$E$-15$  &       ---      & $ 7.48$E$-19$ \\
Na$^+$+H$^-$ & $ 3.67$E$-15$  & $ 5.55$E$-10$  & $ 1.77$E$-07$  & $ 2.41$E$-08$  & $ 7.04$E$-09$  & $ 1.15$E$-11$  & $ 1.75$E$-14$  & $ 1.42$E$-14$  & $ 2.18$E$-17$  &       ---     \\
\multicolumn{11}{c}{\underline{2000 K}} \\
          3s &       ---      & $ 1.71$E$-21$  & $ 7.65$E$-26$  & $ 3.82$E$-27$  & $ 7.85$E$-28$  & $ 7.45$E$-30$  & $ 1.09$E$-31$  & $ 8.46$E$-32$  & $ 2.76$E$-31$  & $ 5.51$E$-27$ \\
          3p & $ 7.86$E$-17$  &       ---      & $ 4.35$E$-15$  & $ 1.51$E$-16$  & $ 4.25$E$-17$  & $ 2.63$E$-20$  & $ 8.94$E$-24$  & $ 8.19$E$-24$  & $ 2.72$E$-25$  & $ 5.87$E$-17$ \\
          4s & $ 8.48$E$-18$  & $ 7.21$E$-12$  &       ---      & $ 7.36$E$-11$  & $ 1.08$E$-11$  & $ 1.95$E$-15$  & $ 1.93$E$-19$  & $ 2.41$E$-19$  & $ 4.63$E$-21$  & $ 2.72$E$-11$ \\
          3d & $ 1.00$E$-18$  & $ 5.91$E$-13$  & $ 1.74$E$-10$  &       ---      & $ 7.55$E$-12$  & $ 1.50$E$-15$  & $ 1.31$E$-17$  & $ 8.22$E$-18$  & $ 1.97$E$-19$  & $ 8.30$E$-12$ \\
          4p & $ 7.57$E$-19$  & $ 6.12$E$-13$  & $ 9.35$E$-11$  & $ 2.77$E$-11$  &       ---      & $ 1.50$E$-11$  & $ 2.08$E$-14$  & $ 2.31$E$-14$  & $ 7.31$E$-17$  & $ 9.04$E$-12$ \\
          5s & $ 1.78$E$-19$  & $ 9.40$E$-15$  & $ 4.19$E$-13$  & $ 1.37$E$-13$  & $ 3.73$E$-10$  &       ---      & $ 3.71$E$-13$  & $ 1.75$E$-13$  & $ 1.67$E$-15$  & $ 4.63$E$-13$ \\
          4d & $ 1.39$E$-21$  & $ 1.69$E$-18$  & $ 2.20$E$-17$  & $ 6.35$E$-16$  & $ 2.74$E$-13$  & $ 1.97$E$-13$  &       ---      & $ 1.90$E$-09$  & $ 3.43$E$-13$  & $ 5.85$E$-16$ \\
          4f & $ 7.94$E$-22$  & $ 1.15$E$-18$  & $ 2.04$E$-17$  & $ 2.94$E$-16$  & $ 2.25$E$-13$  & $ 6.86$E$-14$  & $ 1.41$E$-09$  &       ---      & $ 2.08$E$-13$  & $ 4.75$E$-16$ \\
          5p & $ 8.32$E$-21$  & $ 1.22$E$-19$  & $ 1.26$E$-18$  & $ 2.26$E$-17$  & $ 2.29$E$-15$  & $ 2.10$E$-15$  & $ 8.16$E$-13$  & $ 6.69$E$-13$  &       ---      & $ 7.33$E$-18$ \\
Na$^+$+H$^-$ & $ 2.49$E$-15$  & $ 3.96$E$-10$  & $ 1.11$E$-07$  & $ 1.43$E$-08$  & $ 4.24$E$-09$  & $ 8.74$E$-12$  & $ 2.08$E$-14$  & $ 2.28$E$-14$  & $ 1.10$E$-16$  &       ---     \\
\multicolumn{11}{c}{\underline{4000 K}} \\
          3s &       ---      & $ 2.02$E$-18$  & $ 2.28$E$-21$  & $ 1.43$E$-22$  & $ 3.76$E$-23$  & $ 7.97$E$-25$  & $ 3.28$E$-26$  & $ 3.09$E$-26$  & $ 2.07$E$-25$  & $ 2.61$E$-21$ \\
          3p & $ 3.02$E$-16$  &       ---      & $ 1.51$E$-13$  & $ 1.04$E$-14$  & $ 3.50$E$-15$  & $ 5.73$E$-18$  & $ 4.08$E$-21$  & $ 3.32$E$-21$  & $ 9.77$E$-22$  & $ 4.48$E$-14$ \\
          4s & $ 2.40$E$-17$  & $ 1.07$E$-11$  &       ---      & $ 1.99$E$-10$  & $ 3.79$E$-11$  & $ 1.49$E$-14$  & $ 3.16$E$-18$  & $ 3.29$E$-18$  & $ 4.19$E$-19$  & $ 7.03$E$-10$ \\
          3d & $ 1.04$E$-18$  & $ 5.03$E$-13$  & $ 1.37$E$-10$  &       ---      & $ 9.22$E$-12$  & $ 4.43$E$-14$  & $ 1.11$E$-15$  & $ 6.69$E$-16$  & $ 4.21$E$-17$  & $ 6.05$E$-11$ \\
          4p & $ 6.74$E$-19$  & $ 4.20$E$-13$  & $ 6.45$E$-11$  & $ 2.28$E$-11$  &       ---      & $ 7.51$E$-11$  & $ 2.96$E$-13$  & $ 4.15$E$-13$  & $ 6.83$E$-15$  & $ 4.49$E$-11$ \\
          5s & $ 1.23$E$-19$  & $ 5.93$E$-15$  & $ 2.19$E$-13$  & $ 9.46$E$-13$  & $ 6.48$E$-10$  &       ---      & $ 4.01$E$-12$  & $ 1.79$E$-12$  & $ 3.54$E$-14$  & $ 1.14$E$-12$ \\
          4d & $ 1.65$E$-21$  & $ 1.38$E$-18$  & $ 1.51$E$-17$  & $ 7.73$E$-15$  & $ 8.31$E$-13$  & $ 1.31$E$-12$  &       ---      & $ 2.81$E$-09$  & $ 4.74$E$-12$  & $ 1.53$E$-15$ \\
          4f & $ 1.13$E$-21$  & $ 8.12$E$-19$  & $ 1.14$E$-17$  & $ 3.38$E$-15$  & $ 8.48$E$-13$  & $ 4.24$E$-13$  & $ 2.04$E$-09$  &       ---      & $ 3.00$E$-12$  & $ 1.58$E$-15$ \\
          5p & $ 2.08$E$-20$  & $ 6.56$E$-19$  & $ 3.99$E$-18$  & $ 5.83$E$-16$  & $ 3.82$E$-14$  & $ 2.30$E$-14$  & $ 9.44$E$-12$  & $ 8.21$E$-12$  &       ---      & $ 6.78$E$-17$ \\
Na$^+$+H$^-$ & $ 3.50$E$-15$  & $ 4.03$E$-10$  & $ 8.98$E$-08$  & $ 1.12$E$-08$  & $ 3.37$E$-09$  & $ 9.95$E$-12$  & $ 4.09$E$-14$  & $ 5.79$E$-14$  & $ 9.09$E$-16$  &       ---     \\
\multicolumn{11}{c}{\underline{6000 K}} \\
          3s &       ---      & $ 5.24$E$-17$  & $ 1.25$E$-19$  & $ 5.85$E$-21$  & $ 1.52$E$-21$  & $ 4.13$E$-23$  & $ 4.65$E$-24$  & $ 4.76$E$-24$  & $ 3.03$E$-23$  & $ 3.22$E$-19$ \\
          3p & $ 1.02$E$-15$  &       ---      & $ 5.55$E$-13$  & $ 3.74$E$-14$  & $ 1.27$E$-14$  & $ 2.77$E$-17$  & $ 1.20$E$-19$  & $ 1.72$E$-19$  & $ 4.01$E$-20$  & $ 4.66$E$-13$ \\
          4s & $ 6.02$E$-17$  & $ 1.37$E$-11$  &       ---      & $ 2.37$E$-10$  & $ 4.83$E$-11$  & $ 2.33$E$-14$  & $ 2.33$E$-17$  & $ 2.45$E$-17$  & $ 5.12$E$-18$  & $ 2.03$E$-09$ \\
          3d & $ 1.28$E$-18$  & $ 4.19$E$-13$  & $ 1.08$E$-10$  &       ---      & $ 7.87$E$-12$  & $ 2.28$E$-13$  & $ 6.94$E$-15$  & $ 6.06$E$-15$  & $ 5.61$E$-16$  & $ 1.13$E$-10$ \\
          4p & $ 7.24$E$-19$  & $ 3.09$E$-13$  & $ 4.78$E$-11$  & $ 1.71$E$-11$  &       ---      & $ 1.56$E$-10$  & $ 1.29$E$-12$  & $ 1.40$E$-12$  & $ 5.44$E$-14$  & $ 7.38$E$-11$ \\
          5s & $ 1.19$E$-19$  & $ 4.08$E$-15$  & $ 1.40$E$-13$  & $ 3.00$E$-12$  & $ 9.44$E$-10$  &       ---      & $ 1.11$E$-11$  & $ 5.94$E$-12$  & $ 1.84$E$-13$  & $ 1.69$E$-12$ \\
          4d & $ 3.71$E$-21$  & $ 4.91$E$-18$  & $ 3.87$E$-17$  & $ 2.53$E$-14$  & $ 2.16$E$-12$  & $ 3.08$E$-12$  &       ---      & $ 3.43$E$-09$  & $ 1.68$E$-11$  & $ 3.57$E$-15$ \\
          4f & $ 2.75$E$-21$  & $ 5.07$E$-18$  & $ 2.94$E$-17$  & $ 1.60$E$-14$  & $ 1.70$E$-12$  & $ 1.19$E$-12$  & $ 2.48$E$-09$  &       ---      & $ 1.16$E$-11$  & $ 3.10$E$-15$ \\
          5p & $ 4.54$E$-20$  & $ 3.07$E$-18$  & $ 1.59$E$-17$  & $ 3.84$E$-15$  & $ 1.71$E$-13$  & $ 9.56$E$-14$  & $ 3.14$E$-11$  & $ 3.01$E$-11$  &       ---      & $ 2.91$E$-16$ \\
Na$^+$+H$^-$ & $ 6.22$E$-15$  & $ 4.62$E$-10$  & $ 8.16$E$-08$  & $ 9.94$E$-09$  & $ 3.01$E$-09$  & $ 1.13$E$-11$  & $ 8.66$E$-14$  & $ 1.04$E$-13$  & $ 3.76$E$-15$  &       ---     \\
\multicolumn{11}{c}{\underline{8000 K}} \\
          3s &       ---      & $ 3.77$E$-16$  & $ 1.35$E$-18$  & $ 5.00$E$-20$  & $ 1.25$E$-20$  & $ 4.13$E$-22$  & $ 1.01$E$-22$  & $ 1.03$E$-22$  & $ 5.00$E$-22$  & $ 4.95$E$-18$ \\
          3p & $ 2.66$E$-15$  &       ---      & $ 1.15$E$-12$  & $ 6.84$E$-14$  & $ 2.24$E$-14$  & $ 6.04$E$-17$  & $ 1.13$E$-18$  & $ 2.54$E$-18$  & $ 6.87$E$-19$  & $ 1.66$E$-12$ \\
          4s & $ 1.39$E$-16$  & $ 1.68$E$-11$  &       ---      & $ 2.40$E$-10$  & $ 4.77$E$-11$  & $ 2.80$E$-14$  & $ 1.17$E$-16$  & $ 1.79$E$-16$  & $ 4.34$E$-17$  & $ 3.43$E$-09$ \\
          3d & $ 1.90$E$-18$  & $ 3.69$E$-13$  & $ 8.88$E$-11$  &       ---      & $ 7.07$E$-12$  & $ 6.16$E$-13$  & $ 1.84$E$-14$  & $ 2.49$E$-14$  & $ 3.13$E$-15$  & $ 1.51$E$-10$ \\
          4p & $ 9.70$E$-19$  & $ 2.46$E$-13$  & $ 3.60$E$-11$  & $ 1.44$E$-11$  &       ---      & $ 2.41$E$-10$  & $ 3.05$E$-12$  & $ 2.92$E$-12$  & $ 1.91$E$-13$  & $ 9.36$E$-11$ \\
          5s & $ 1.62$E$-19$  & $ 3.36$E$-15$  & $ 1.07$E$-13$  & $ 6.36$E$-12$  & $ 1.23$E$-09$  &       ---      & $ 2.02$E$-11$  & $ 1.28$E$-11$  & $ 6.63$E$-13$  & $ 2.13$E$-12$ \\
          4d & $ 1.02$E$-20$  & $ 1.61$E$-17$  & $ 1.14$E$-16$  & $ 4.86$E$-14$  & $ 3.96$E$-12$  & $ 5.15$E$-12$  &       ---      & $ 3.91$E$-09$  & $ 3.91$E$-11$  & $ 6.39$E$-15$ \\
          4f & $ 7.43$E$-21$  & $ 2.60$E$-17$  & $ 1.26$E$-16$  & $ 4.74$E$-14$  & $ 2.73$E$-12$  & $ 2.36$E$-12$  & $ 2.82$E$-09$  &       ---      & $ 2.72$E$-11$  & $ 5.04$E$-15$ \\
          5p & $ 9.13$E$-20$  & $ 1.78$E$-17$  & $ 7.73$E$-17$  & $ 1.50$E$-14$  & $ 4.51$E$-13$  & $ 3.08$E$-13$  & $ 7.12$E$-11$  & $ 6.88$E$-11$  &       ---      & $ 8.48$E$-16$ \\
Na$^+$+H$^-$ & $ 1.15$E$-14$  & $ 5.46$E$-10$  & $ 7.75$E$-08$  & $ 9.22$E$-09$  & $ 2.81$E$-09$  & $ 1.26$E$-11$  & $ 1.48$E$-13$  & $ 1.62$E$-13$  & $ 1.08$E$-14$  &       ---     \\
\hline
\end{tabular}
\end{center}
\end{table*}

The accuracy of these rate coefficients is determined by the accuracy of the cross-sections near the threshold.  At $T=2000$~K the values of the rate coefficients vary from $10^{-31}$ to $10^{-9}$~cm$^3/$s, more than 20 orders of magnitude.  At $T=8000$~K the variation is from $10^{-22}$ to $10^{-9}$~cm$^3/$s, 13 orders of magnitude. This reflects the fact that at the collision energies of interest the nonadiabatic transition probabilities also vary by similar factors, and that many of them have extremely small values. These small transition probabilities are sensitive to variations in both the input quantum chemical data (potentials and couplings) and the method employed for solving the coupled channel equations in the quantum scattering calculations.\footnote{Note, in \cite{2010PhRvA..81c2706B} we found satisfactory agreement between two different methods and codes when calculating cross-sections for the $3s\rightarrow 3p$ and $4s$ transitions.   These transitions have reasonably large cross-sections.  However, recent additional calculations have shown some differences between calculations for transitions with much smaller cross-sections.}  In addition, transitions can occur in symmetries other than $^1 \Sigma^+$: for example, in the singlet system the $p$-, $d$-, and $f$-states can populate the ionic state via rotational couplings between $\Pi$ and $\Sigma$ states, as well as excite other states due to both rotational and radial nonadiabatic couplings in $\Pi$, $\Delta$, and other symmetries.  The same is true for the triplet system with larger (at least by a factor of 3) initial statistical probabilities.  For example, the cross-sections for $3s \rightarrow 3p$ excitation and their sensitivity to the input quantum chemical data are discussed at length in \cite{2010PhRvA..81c2706B}, where we have shown that the results may vary by a factor of about 100.   Similarly, from our investigations it is possible to make estimates of how much rate coefficients might vary for all transitions, and thus we define a ``fluctuation factor'' representing an estimate of this variation and which is somewhat analogous to an estimate of the uncertainty.  
The probable range for a given rate coefficient can be estimated to be from the value given in Table~\ref{tab:rates} multiplied by the minimum value of the fluctuation factor, up to the same value multiplied by the maximum value of the fluctuation factor, noting that from the arguments discussed above, particularly contributions from other symmetries, it is generally more likely that the rate coefficient would be increased rather than decreased.   

We have assembled estimates of these fluctuation factors for all transitions in Table~\ref{tab:errors}.  A common feature of the quantum scattering calculations, particularly for very adiabatic transitions, is that the sensitivity of the cross-sections to the input quantum chemistry data usually decreases with increasing collision energy, and thus cross-sections are often most uncertain very close to threshold, becoming less uncertain with increasing collision energy.  This means that rate coefficients are usually most uncertain at low temperatures, becoming less uncertain at higher temperatures. Thus, in some cases we provide a range of fluctuation factors corresponding to the lower and higher ends of the temperature range considered.  The rate coefficient for $3s \rightarrow 3p$ excitation is roughly $10^{-21}$~cm$^3/$s at 2000~K, somewhere in the middle of the range for all transitions.   Thus, it is quite natural to expect that fluctuation factors for transitions with small transition probabilties can be much larger than 100, while the fluctuation coefficients for the transitions determined by large transition probabilities should be close to unity. 

It is worth making clear that the smallest rate coefficients usually have the largest fluctuation factors: even if these rates were multiplied by such a large fluctuation factor they would not become dominant or substantially change the general pattern of relative strengths between different transitions.  Nevertheless, for astrophysical applications it might be important to have such estimates of the possible uncertainties for the rate coefficients in order to see if any such transitions could have effects and thus warrant more accurate study.

\begin{table*}
\caption{Estimated \emph{maximum} values of the ``fluctuation factors'' (a measure of the uncertainty) in the rate coefficients, given as factors of the rate coefficients in Table~\ref{tab:rates}.  The minimum value is estimated to be 0.5 in all cases.  If a range is given, the first number corresponds to 2000~K and the second the 8000~K temperature result.  For example, if the table lists 10--2, then the maximum fluctuation factor in the rate coefficient is estimated to be 10 near 2000~K and 2 near 8000~K.} 
\label{tab:errors}
%\scriptsize
\tiny
\begin{center}
\begin{tabular}{ccccccccccc}
\hline
Initial & \multicolumn{8}{c}{final state $n^\prime l^\prime$}  \\
state $nl$   & 3p & 4s & 3d & 4p & 5s & 4d & 4f & 5p & Na$^+$+H$^-$  \\
\hline
3s           & 120--70 & 50--10 & 10 & 20 & 100--10 & 100--10 & 100--10 & $10^3$--100 & 2 \\ 
3p           &         & 2 & 5 & 2 & 5--3 & 100--10 & $10^3$--30 & $10^5$--100  & 2\\
4s           &         &   & 6 & 2 & 100--10 & $10^4$--400 & $10^4$--100 & $10^5$--$10^3$ & 2\\
3d           & & & & 2 & 20--2 & 100--20 & 10--2 & $10^4$--2 & 2--3 \\
4p           & & & & & 2 & 2 & 2 & 2 & 2 \\
5s           & & & & & & 2 & 2 & 100--2 & 10 \\
4d           & & & & & & & 2 & $10^3$--2 & $10^4$--100 \\
4f           & & & & & & & & 10--2 & 100--40 \\
5p           & & & & & & & & & $10^4$--30 \\
\hline
\end{tabular}
\end{center}
\end{table*}

These calculations have been employed to determine the total rate coefficient for mutual neutralization. This rate coefficient can be approximated as $2.44 \times 10^{-7} (300\mathrm{K}/T)^{0.32}$~cm$^3/$s with an error of less than 5\% for $500 \le T  \le 8000$~K. This new result can be up to 50\% larger than that given by \cite{DPG:99}: the difference is primarily because a very much finer energy grid has been employed in the \cite{2010PhRvA..81c2706B} work.

\section{Discussion}

\begin{figure}
\begin{center}
\resizebox{90mm}{!}{\includegraphics{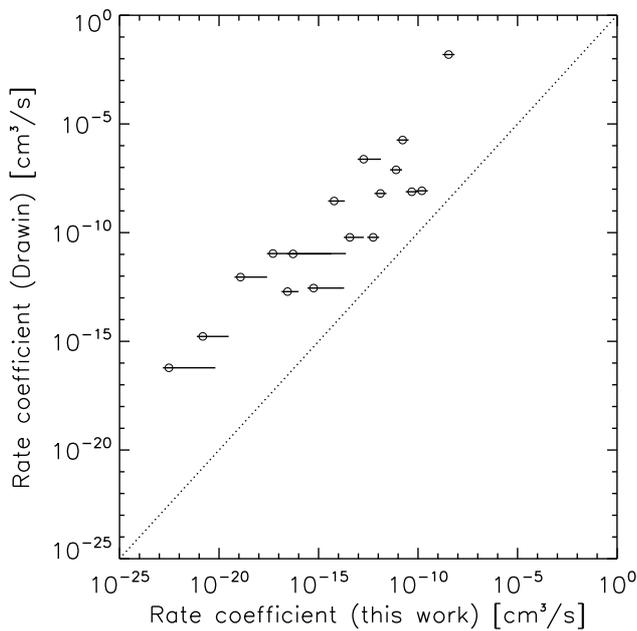}}
\end{center}
\caption{Comparison of rate coefficients at 6000~K from the Drawin formula (y-axis) with those from this work (x-axis).  The dotted line is the one-to-one relation.  The horizontal lines show the range in our values implied by the fluctuation factors, which are taken from Table~\ref{tab:errors} and log-interpolated at 6000~K between the extremes if a range is given.}
\label{fig:1}
\end{figure}

The rate coefficients show a general pattern similar to that found for Li+H collisions \citep{2003A&A...409L...1B}, with large rates for charge exchange processes, particularly from the first excited $s$ state, and comparatively small rates for excitation processes, except in the cases of transitions between neighbouring excited levels.  One may therefore reasonably expect similar effects on the statistical equilibrium calculations from inclusion of these processses: i.e.\, excitation and deexcitation processes have little or negligible effect, while charge exchange processes have significant effects.   It is, however, important that this be confirmed via detailed statistical equilibrium calculations.  

Regarding comparison of these data with those from the commonly used Drawin formula \citep{1984A&A...130..319S,1968ZPhy..211..404D,1973PhLA...43..333D}, the situation is again similar to that for Li.   The Drawin formula gives rate coefficients for optically allowed transitions (the Drawin formula cannot be used for optically forbidden transitions or charge exchange) between one and seven orders of magnitude larger than the results presented here, see Fig.~\ref{fig:1}.  Note that there is some correlation between the Drawin results and ours which simply stems from the fact that the Drawin formula correctly accounts for the fact that transitions with smaller energy thresholds will have larger rate coefficients.  In any case there is a scatter of some six orders of magnitude around this correlation and a significant offset.  We also note that the differences cannot be explained within the estimated fluctuation factors for the results, and in addition there is no correlation between the difference and the uncertainty in our rate coefficient.

\begin{acknowledgements}
We gratefully acknowledge the support of the Royal Swedish Academy of Sciences, G{\"o}ran Gustafssons Stiftelse and the Swedish Research Council.  P.S.B is a Royal Swedish Academy of Sciences Research Fellow supported by a grant from the Knut and Alice Wallenberg Foundation.   A.K.B. gratefully acknowledges the support from the Russian Foundation for Basic Research (Grant No. 10-03-00807-a).
\end{acknowledgements}

\bibliographystyle{aa} 
\bibliography{15152a,15152b,15152c,15152d,15152e}

\end{document}